\documentclass[preprint]{aastex}

\newcommand{\jcpp}{J. Comp. Phys.}

\begin{document}

\title{
Instability of Non-uniform Toroidal Magnetic Fields in Accretion Disks
}

\author{Kota Hirabayashi and Masahiro Hoshino}
\affil{Department of Earth and Planetary Science, The University of
Tokyo, 7-3-1, Hongo, Bunkyo-ku, Tokyo 113-0033, Japan}
\email{hirabayashi-k@eps.s.u-tokyo.ac.jp}

 \begin{abstract}
  A new type of instability that is expected to drive
  magnetohydrodynamic (MHD) turbulence from a purely toroidal magnetic
  field in an accretion disk is presented.
  It is already known that in a differentially rotating system, the
  uniform toroidal magnetic field is unstable due to a magnetorotational
  instability (MRI) under a non-axisymmetric and vertical perturbation,
  while it is stable under a purely vertical perturbation.
  Contrary to the previous study, this paper proposes an unstable mode
  completely confined to the equatorial plane, driven by the expansive
  nature of the magnetic pressure gradient force under a non-uniform
  toroidal field.
  The basic nature of this growing eigenmode,
  to which we give a name ``magneto-gradient driven instability'',
  is studied using linear analysis, and the corresponding nonlinear
  evolution is then investigated using two-dimensional ideal MHD
  simulations.
  Although a single localized magnetic field channel alone cannot
  provide sufficient Maxwell stress to contribute significantly to the
  angular momentum transport, we find that the mode coupling between
  neighboring toroidal fields under multiple localized magnetic field
  channels drastically generates a highly turbulent state and leads to
  the enhanced transport of angular momentum, comparable to the
  efficiency seen in previous studies on MRIs.
  This horizontally confined mode may play an important role in the
  saturation of an MRI through
  complementray growth with the toroidal MHDs and
  coupling with magnetic reconnection.
 \end{abstract}

 \keywords{accretion, accretion disks --- instabilities ---
 magnetohydrodynamics --- methods: numerical --- turbulence}
 
 \section{INTRODUCTION}
 Accretion disks are one of the most ubiquitous astrophysical objects,
 comprising dynamics such as astrophysical jets, disk winds, and
 particle acceleration.
 It is widely believed that these dynamical phenomena are driven by the
 anomalous transport of angular momentum and the subsequent release
 of gravitational binding energy. 
 Several mechanisms have been proposed in attempts to explain the origin
 of this angular momentum transport.
 Examples include magnetic braking by external, large-scale magnetic
 fields \citep[e.g.,][]{Blandford1982,Stone1994}, non-axisymmetric wave
 excitation \citep[e.g.,][]{Fragile2008}, and hydrodynamic/hydromagnetic
 turbulence \citep[e.g.,][]{Papaloizou1984,Balbus1998}.
 In this paper, we consider a mechanism related to the third example,
 which is the sole candidate that possesses a high correlation with
 conventional, $\alpha$-disk models \citep{Shakura1973}.
 In the $\alpha$-disk models, the efficiency of angular momentum
 transport, which is determined by the $R\phi$-component of the stress
 tensor, is determined as the product of the pressure and a given
 parameter $\alpha$.
 The value of $\alpha$ depends significantly on viscosity physics, but
 the simple molecular viscosity in an accretion disk cannot provide a
 high efficiency of angular momentum transport suggested by observations
 \citep{Cannizzo1988}.
 Since the astrophysical importance of magnetorotational instability
 (MRI) as the origin of required turbulence was pointed out
 \citep{Balbus1991,Balbus1998}, a number of authors have investigated
 the nature of MRIs and the resultant turbulence in accretion disks over
 a wide range of plasma parameters
 \citep[e.g.,][]{Stone1996,Sano2002,Kunz2013,Hoshino2015,Bai2015,Zhu2015,Simon2009,Simon2012}.

 In order to study the basic behavior behind the non-linear time
 evolution of MRIs, most numerical studies on the local properties of
 MRI-induced turbulence have adopted the shearing box model
 \citep{Hawley1995,Sano2001,Sharma2006}, which can capture the wave
 vector toward an arbitrary direction in a differentially rotating
 plasma.
 Since an MRI with a vertical wave vector has the maximum growth rate
 for an axisymmetric perturbation when the background magnetic field is
 purely poloidal, fully three-dimensional simulations, or at least
 two-dimensional ones including a vertical axis, are necessary.
 (Note that the final states in two- and three-dimensional cases are
 rather different from each other, and that the three-dimensional
 simulations are required to investigate the saturation stage.)

 The situation is similar when the unperturbed magnetic field is purely
 toroidal.
 For exapmle, \cite{Balbus1992} investigated the linear stability of an
 accretion disk threaded by a uniform toroidal magnetic field assuming
 three-dimensional wavevectors in the cylindrical coordinates, whose
 $x$-component varies with time because of the background shear
 velocity.
 They showed that the perturbation satisfying
 ${\bf k}\cdot{\bf V}_A \lesssim \Omega$ can become unstable, in the
 sense that the amplitude of oscillation increases with time.
 Moreover, a finite vertical wavenumber, $k_z$, is required for the
 instability to occur, and the larger $k_z$ leads to the faster
 amplification.
 The non-linear evolution of this oscillatory instability was also
 examined by \cite{Hawley1995} using three-dimensional ideal MHD
 simulations, and the contribution to turbulence generation was
 confirmed.

 Other examples include linear eigenvalue analyses and the
 corresponding MHD simulations by \cite{Matsumoto1995}.
 They revealed that purely growing eigenmodes can exist in a shearing
 plasma, in contrast to the above oscillatory unstable modes.
 For a Keplerian disk, only the non-axisymmetric perturbations with
 $k_y^2/k_z^2 < 0.015$ become purely growing modes, where $k_y$ and
 $k_z$ are the azimuthal and vertical wavenumbers, respectively.
 The vertical waves, therefore, again contribute to the unstable modes
 most significantly, although a finite azimuthal wavenumber is
 required.

 Such a situation, where the toroidal magnetic field is dominant, is
 thought to appear easily in the non-linear stage of an MRI even when
 starting from a poloidal field.
 For understanding the dynamics and the nature of turbulence in
 well-developed disks, therefore, it is important to investigate a
 plasma stability under a purely toroidal field.
 In this paper, we provide the results of linear and non-linear analyses
 on this issue, focusing particularly on an initially non-uniform
 toroidal field, and suggest another possible path leading to turbulence
 generation.
 Unstable modes proposed here, which we will call ``magneto-gradient
 driven instability (MGDI)`` reflecting its driving source as shown in
 the suceeding sections, are confined within the equatorial plane,
 i.e., $k_z=0$, in contrast to the previous studies that required
 $k_z^2>k_y^2$.
 An instability bound to the equatorial plane may play a crucial
 role in plasma transport, as it could potentially couple with magnetic
 reconnection occurring in the plane and contribute to the saturation
 mechanism of MRIs.

 The outline of this paper is as follows.
 In Section \ref{sec:linear}, we briefly introduce the setup of our
 theoretical study and show the existence of unstable eigenmodes by
 linear analysis.
 Section \ref{sec:nonlinear} discusses the results of the
 fully-non-linear two-dimensional numerical simulations.
 The non-linear calculations corresponding to the linear study and ones
 which can lead to more turbulent states are presented.
 Finally, Section \ref{sec:summary} is devoted to the summary and
 conclusion of our results.

 \section{LINEAR ANALYSIS}\label{sec:linear}
 In this section we investigate the linear stability of a non-uniform
 toroidal magnetic field in a differentially rotating system.
 In particularly, a simplified situation with a localized toroidal
 magnetic field channel is considered to extract the physical essence of
 possible unstable modes.

  \subsection{Equilibrium State and Linearized Equations}
  The ideal MHD equations incorporated with the standard 
  shearing box model are employed as the basic equations
  throughout this paper \citep{Stone2010}:
  \begin{eqnarray}
   \frac{\partial \rho}{\partial t}
    + {\bf v} \cdot \nabla \rho &=& 
     - \rho \nabla \cdot {\bf v},
     \label{eq:linear1} \\
   \rho \left(\frac{\partial {\bf v}}{\partial t}
	 + {\bf v} \cdot \nabla {\bf v} \right)
   &=&
   - \nabla \left( p + \frac{B^2}{2} \right)
   + {\bf B} \cdot \nabla {\bf B}
   - 2 \rho {\bf \Omega} \times {\bf v}
    - 2 \rho \Omega x v_{y0}^{\prime} \hat{\bf e}_x,
   \label{eq:linear2} \\
   \frac{\partial {\bf B}}{\partial t}
    &=&
    \nabla \times \left( {\bf v} \times {\bf B} \right),
    \label{eq:linear3} \\
   \frac{\partial p}{\partial t}
    + {\bf v} \cdot \nabla p
    &=& - \gamma p \nabla \cdot {\bf v}.
    \label{eq:linear4}
  \end{eqnarray}
  The radial and azimuthal directions are then interpreted as the $x$-
  and $y$-axes in the local Cartesian coordinate system, where the
  differential rotation is described by a linearly changing background
  velocity defined as $v_{y0}\left(x\right) = -q \Omega x$, using an
  angular velocity at the center of the computational domain, $\Omega$,
  and a positive constant, $q$.
  The specific heat ratio, $\gamma$, is set to be 5/3, and the other
  notations are standard.

  When a purely toroidal magnetic field is imposed, the background
  shearing plasma is kept in an equilibrium state as long as the total
  pressure is spatially constant.
  We can, therefore, choose an arbitrary magnetic structure with a
  finite gradient.
  Here, we focus on the idealized case with a simple localized toroidal
  field,
  \begin{equation}
   B_{y0}\left(x\right) = B_0 \cosh^{-2} \left(x/d\right),
  \end{equation}
  where $B_0$ is the field strength at $x=0$
  and $d$ is the typical width of the localized field.
  The gas pressure is determined so as to satisfy the total pressure
  balance, and the background density is distributed so as to keep the
  temperature uniform.

  Next, we linearize the MHD equations around this equilibrium state,
  assuming the functional form of a small perturbation as
  \begin{equation}
   f_1 \left(x,y,t\right) =
    f_1\left(x\right)
    \exp \left(-i\omega t + ik_y y\right).
  \end{equation}
  The vertical dependence is ignored so as to pick up
  horizontally confined modes.
  Using the vector potential instead of the magnetic field, the
  linearization of the basic equations leads to the following eigenvalue
  problem:
    \begin{equation}
   \omega {\bf U}_1 = {\bf M} {\bf U}_1,
    \label{eq:eigen}
  \end{equation}
  where
  ${\bf U}_1 = (v_{x1}\;\;v_{y1}\;\;A_{z1}\;\;p_1)^T$
  is a first-order perturbation vector, and ${\bf M}$ is a coefficient
  matrix whose components are given as follows,
  \begin{eqnarray*}
   {\bf M} = k_y v_{y0} {\bf I} + {\bf D},
  \end{eqnarray*}
  \begin{eqnarray*}
   {\bf D} =
    \left(
     \begin{array}{cccc}
      0 & 2i\Omega &
       D_{13} &
       - \left(i/\rho_0\right) \partial_x \\
      D_{21} & 0 &
       -k_y B_{y0}^{\prime}/\rho_0 & k_y/\rho_0 \\
      iB_{y0} & 0 & 0 & 0 \\
      D_{41} & \gamma k_y p_0 & 0 & 0
     \end{array}
    \right),
  \end{eqnarray*}
  \begin{eqnarray*}
   D_{13} &=&
    \left(i/\rho_0\right)
    \left[ B_{y0}\left(\partial_x^2-k_y^2\right)
     + B_{y0}^{\prime} \partial_x \right], \\
   D_{21} &=& -iv_{y0}^{\prime}-2i\Omega, \\
   D_{41} &=& -ip_0^{\prime}-i\gamma p_0\partial_x.
  \end{eqnarray*}
  Here, ${\bf I}$ is the identity matrix, and $\partial_x$ and
  $\cdot^{\prime}$ indicate differentiation operators by $x$.
  Note that the equations for $v_{z1}$ and $B_{z1}$ are
  decoupled as usual shear Alfv\'{e}n waves.
  Although the density perturbation, $\rho_1$, appears to have
  disappeared from the basic equations, compressional modes remain in
  the system and $\rho_1$ can be obtained from $\nabla\cdot{\bf v}_1$.

  Finally, equation (\ref{eq:eigen}) is discretized in the computational
  domain $\left|x/L_x\right| \le 1$ with 400 grid points using a
  fourth-order central difference.
  The width of the localized field is set to be $d=0.05L_x$.
  As a boundary condition, a conducting wall is assumed at
  $\left|x/L_x\right|=1$.
  The eigenvalues, $\omega$, and the eigenvectors, ${\bf U}_1$, are
  then computed numerically.

  \subsection{Growth Rates}
\begin{figure}[htbp]
   \epsscale{0.5}
   \plotone{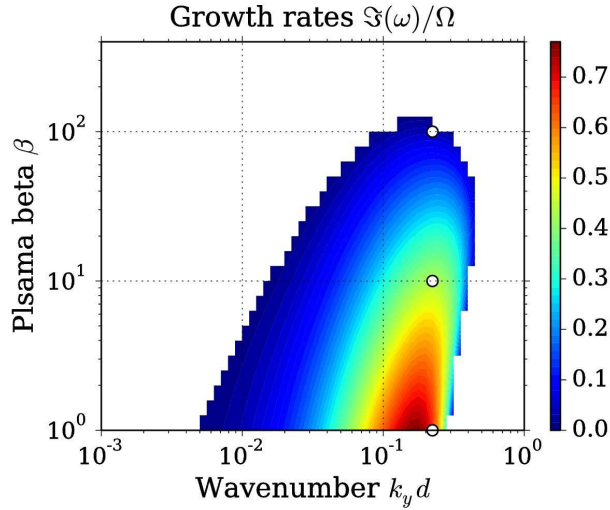}
   \caption{The color contour of growth rates as a function of the
   plasma beta and the wavenumber normalized by $d$, which is the width
   of the localized toroidal field. The gradient of the angular velocity,
   $q$, is set to unity.
   \label{fig:1}}
  \end{figure}

  From our calculations of the eigenvalue problem described above, we
  obtained one growing mode for the MGDI at most for each
  particular wavenumber value.
  The results are summarized in Figure \ref{fig:1}, where the color
  contour shows the imaginary parts of the eigenvalues as a function of
  the wavenumber, normalized by the width of the localized field, and the
  plasma beta, $\beta=2p/B^2$, measured at $x=0$.
  Note that the real parts are zero in a machine precision.

  Figure \ref{fig:1} indicates that the purely growing mode appears if
  $\beta$ is lower than about 100, and that the growth rate becomes
  larger as the initial magnetic field strength increases.
  When $\beta$ is equal to unity, the maximum growth rate reaches
  $0.765\Omega$, which is comparable to that of the axisymmetric MRI,
  i.e., $0.75\Omega$, as far as the linear approximation is appropriate.
  We should emphasize here that the magnitude of the velocity shear,
  $q$, is assumed to be unity in Figure \ref{fig:1} for theoretical
  simplicity, which is smaller than in the case of Keplerian rotation,
  where $q=1.5$.
  In the Keplerian rotation case, we expect more unstable eigenmodes due
  to the stronger shear motion.

  A physical picture of the MGDI can be explaind as follows.
  Let us consider an outward going perturbation in $v_{x1}$ away
  from $x=0$, i.e., positive for $x>0$ and negative for $x<0$.
  Since a linearized equation for the azimuthal magnetic field can be
  written as 
  \begin{eqnarray}
   \frac{dB_{1y}}{dt} =
    - B_{y0} \frac{\partial v_{x1}}{\partial x}
    - B_{y1}^{\prime} v_{x1}
    - q \Omega B_{x1},
    \label{eq:linear-by}
  \end{eqnarray}
  where $d/dt=\partial/\partial t + {\bf v}_0 \cdot \nabla$ is a
  Lagrangian derivative,
  such outward $v_{x1}$ directly induces the increment in $B_{y1}$
  through the second term in the right-hand side.
  This term comes from the linearized advection effect, which represents
  the fact that a fluid element brings the frozen-in magnetic field line
  from the original position with the stronger magnetic field.
  On the other hand, a lienarized version of the equation of motion for
  the radial velocity is as follows,
  \begin{eqnarray}
   \rho_0 \frac{dv_{x1}}{dt} = 
    - \frac{\partial p_1}{\partial x}
    - \frac{\partial}{\partial x}\left( B_{y0} B_{y1} \right)
    + B_{y0} \frac{\partial B_{x1}}{\partial y}
    + 2 \rho_0 \Omega v_{y1}.
    \label{eq:lienar-vx}
  \end{eqnarray}
  The second and third terms in the right-hand side represent the
  magnetic pressure gradient and the magnetic tension force,
  respectively.
  The magentic pressure can be further decomposed into two contribution
  from $-B_{y0}^{\prime} B_{y1}$ and $-B_{y0} \partial_x B_{y1}$.
  The increase in $B_{y1}$, then, leads to further expansion force via
  the first component of the magnetic pressure, which implies a positive
  feedback.
  This feedback process will continue to work as long as the finite
  gradient in the background magnetic field is available.

\begin{figure}[htbp]
   \epsscale{0.5}
   \plotone{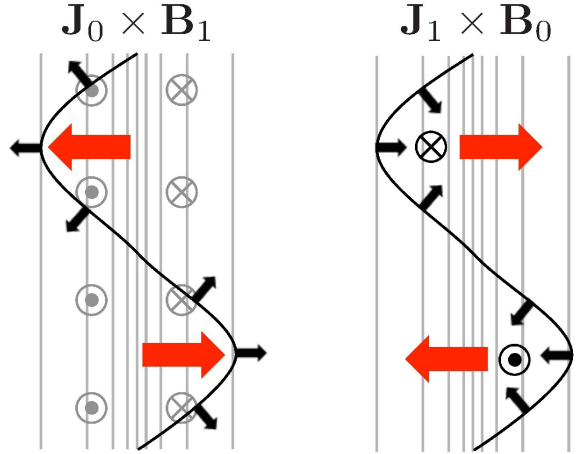}
   \caption{Schematic view of first-order Lorentz forces.
   In the left panel, the background current ($\mathbf{J}_0$; gray
   circles) makes expansive Lorentz force (black arrows) combined with
   the perturbed magnetic field ($\mathbf{B}_1$; black lines).
   The resultant force works to increase the purturbation, like shown by
   the red arrows.
   In the right panel, on the other hand, a cross product of the
   first-order current ($\mathbf{J}_1$; black circles), which is
   generated by the perturbed magnetic field, and the background magnetic
   field ($\mathbf{B}_0$; gray lines) makes the inward Lorentz force,
   which always works as the restoreing force.
   \label{fig:Force}}
  \end{figure}

  As well as the growth rates, the range of unstable wavenumbers also
  tends to broaden as the plasma beta decreases, especially toward the
  long-wavelength side.
  The smallest scale, on the other hand, seems to always be limited
  roughly by $k_y d < 0.5$, which corresponds to the wavelength one
  order of magnitude larger than $d$.
  This bound could be understood qualitatively by the competition
  between the magentic pressure gradient force, which is a driver here,
  and the magnetic tension force working as a restoring force.
  For the feedback mechanism described above to occur, it is clear that
  the expansive nature of the magnetic pressure needs to be greater than
  the tension effect.
  These promoting and restoring effetcs can be rearranged into the form
  of the Lorentz force, ${\bf J}_1 \times {\bf B}_0$ and
  ${\bf J}_0 \times {\bf B}_1$.
  The schematic view of the situation is illustrated in
  Fig.\ref{fig:Force}.
  Thenm the condition that the expansive term outpaces the restoring force is
  roughly estimated as
  \begin{eqnarray*}
   \left| \frac{\partial B_{y0}}{\partial x} B_{y1} \right| > 
    \left| \frac{\partial B_{x1}}{\partial y} B_{y0} \right|.
  \end{eqnarray*}
  Replacing the derivatives by typical scales like $1/d$ and $k_y$, we
  obtain the estimate, $1/d > k_y$, which is consistent with the
  unstable range in Figure \ref{fig:1}.
 
\begin{figure}[htbp]
   \epsscale{0.5}
   \plotone{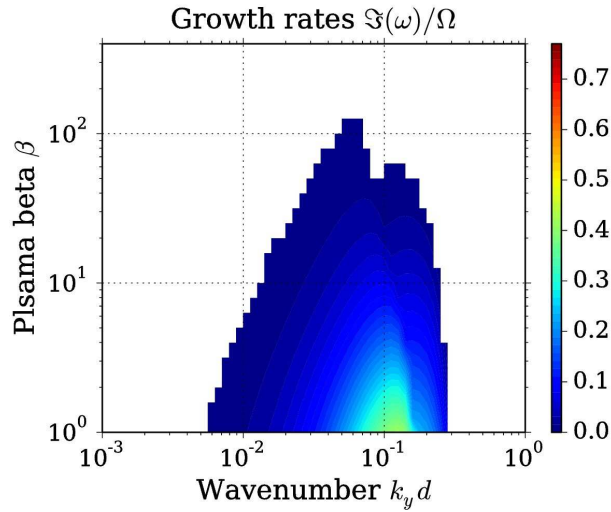}
   \caption{The color contour of growth rates in a rigid-rotating plasma,
   $q=0$, with the same format as in Figure \ref{fig:1}.
   \label{fig:2}}
  \end{figure}

  The destabilization mechanism described above seems not to be related
  to the nature of the differential rotation.
  For the purpose of comparison, growth rates in a rigid-rotating plasma
  are shown in Figure \ref{fig:2} with the same format as in Figure
  \ref{fig:1}.
  This panel indicates the existence of unstable modes over a similar
  range to that in the differentially-rotating case.
  The qualitative dependence on $k_y$ and $\beta$ also resembles that in
  Figure \ref{fig:1}, but the magnitude of the growth rate becomes
  smaller by a factor of 2 for $\beta=1$, and much more for larger
  $\beta$.
  Therefore, it can be concluded that this instability arises originally
  from the gradient of the magnetic field, and can attain a large growth
  rate comparable to that of the standard MRI only when combined with
  the shearing motion.

  \subsection{Eigenfunctions}
\begin{figure*}[htbp]
   \epsscale{0.8}
   \plotone{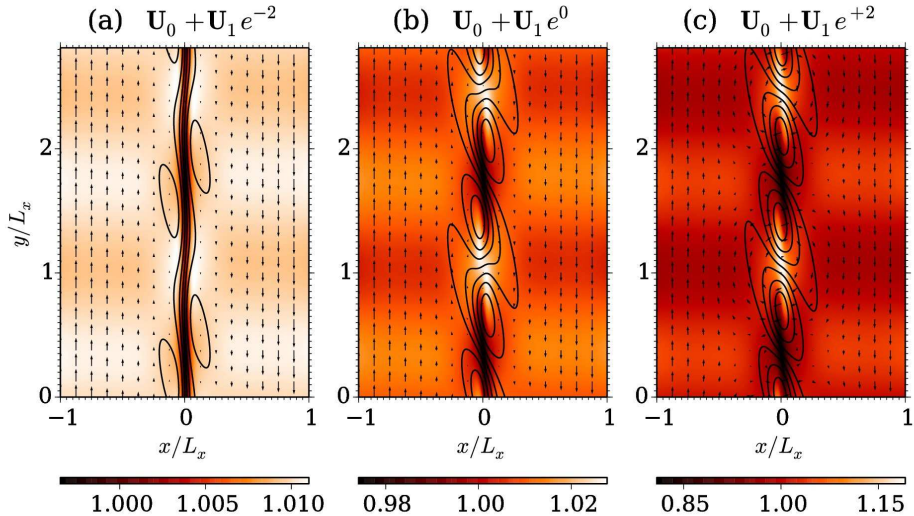}
   \caption{Two-dimensional views of an unstable eigenfunction with the
   wavenumber $k_yd=0.223$, superposed on the background plasma with
   $\beta=100$.
   The corresponding growth rate is $0.039\Omega$.
   The color contour, the solid lines, and the vector map indicate the
   gas pressure, the magnetic field lines, and the velocity,
   respectively.
   The first-order eigenfunction ${\bf U}_1$ is normalized by
   $\left|{\bf U}_1\right|=1$.
   From left to right, the amplitude of the eigenfunction increases
   $2e$-fold.
   \label{fig:3}}
  \end{figure*}

  Let us discuss the structure of the eigenfunctions.
  Figure \ref{fig:3} shows two-dimensional representations of the
  Fourier decomposed eigenfunction, ${\bf U}_1$, superposed on the
  background equilibrium state, ${\bf U}_0$, where ${\bf U}_1$ is
  normalized to satisfy $\left|{\bf U}_1\right|=1$.
  Based on the normalized case in panel (b), the states before and
  after twice the $e$-folding time are shown in panels (a) and (c),
  respectively.
  The plasma beta and the wavenumber are chosen to be $\beta=100$ and
  $k_yd$=0.223, respectively, and the corresponding point on the
  $k_y$-$\beta$ diagram is plotted in Figure \ref{fig:1} by an outlined
  circle.
  In each panel, the color contour, the solid lines, and the vector
  field show the gas pressure distribution, the lines of magnetic force,
  and the bulk velocity, respectively.
 
  Figure \ref{fig:3} shows that bending of the field line broadens with
  time, and eventually spreads out beyond the typical width of the
  equilibrium field, $d=0.05L_x$.
  This broadening of the field line is explained by the destabilization
  process described in the previous subsection, i.e., the outward
  magnetic pressure gradient exceeding the inward magnetic tension
  force may further expand the magnetic field explosively.
  In addition to the expansion in the $x$-direction, the field lines are
  also stretched in the $y$-direction by the so-called $\Omega$-effect
  due to the background shear motion, and thus the magnetic field lines
  become inclined downward to the right.
  In other words, $B_x$ and $B_y$ tend to have negative correlation.
  Therefore, it can be expected that the MGDIs have the potential
  to contribute to powerful angular momentum transport, making the
  averaged Maxwell stress, $\left<-B_x B_y\right>$, once developing to
  non-linear turbulence.

  Other important features include vortex structures around the nodes of
  the magnetic field lines in panel (c).
  In particular, the clockwise vortices at every other node, which align
  with the differential rotation, are selectively enhanced.
  Since the Coriolis force works rightward to the direction of motion,
  the inside of the clockwise vortex is compressed, and the other is
  expanded.
  On the other hand, the selective enhancement makes the magnetic field
  lines loosened around the clockwise vortex, as if a tightly stretched
  rope is reeled up.
  This leads to the negative correlation between the gas pressure and
  the magnetic pressure, which implies that the present unstable mode
  essentially arises from slow-magnetosonic waves.

  Looking at Figure \ref{fig:3}, one may associate the MGDIs with
  current-driven instabilities (CDIs), which are thought to contribute
  fast dissipation of magnetic energy in various astrophysical contexts
  \citep[e.g.,][]{Mignone2010,ONeill2012,Mizuno2014}.
  Although the characteristic that both unstable modes are driven by
  magnetic non-uniformity is common, we consider the MGDI as a
  fundamentally different mode from the CDI.
  To make the difference clear, for example, notice that the wavevector
  in the CDI is essentially parallel to the background electric current,
  which drives the instability.
  In the case of the MGDI, on the other hand, the background current has
  only an out-of-plane component, which is obviously perpendicular to
  wavevector of the perturbation.

\begin{figure*}[htbp]
   \epsscale{0.8}
   \plotone{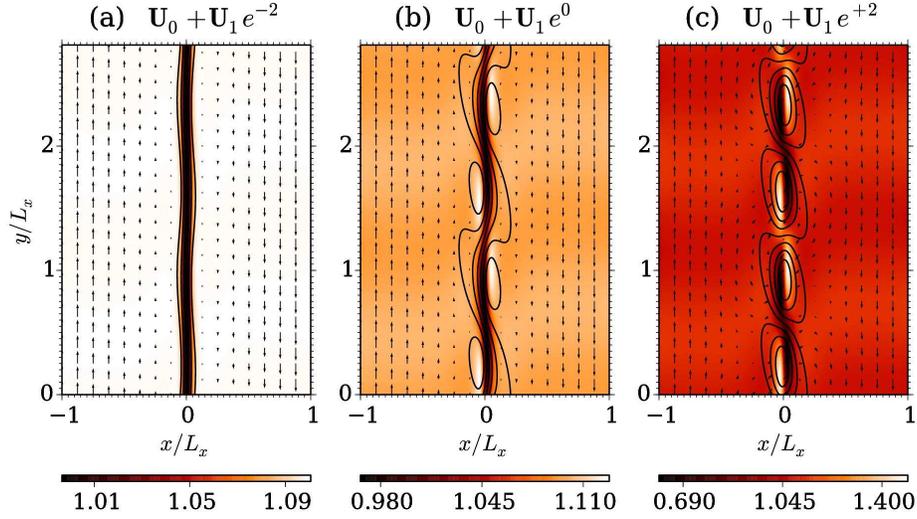}
   \caption{Unstable eigenfunction with $\beta=10$ and $k_yd=0.223$.
   The format is same as in Figure \ref{fig:3}.
   The growth rate is $0.430\Omega$.
   \label{fig:4}}
  \end{figure*}

\begin{figure*}[htbp]
   \epsscale{0.8}
   \plotone{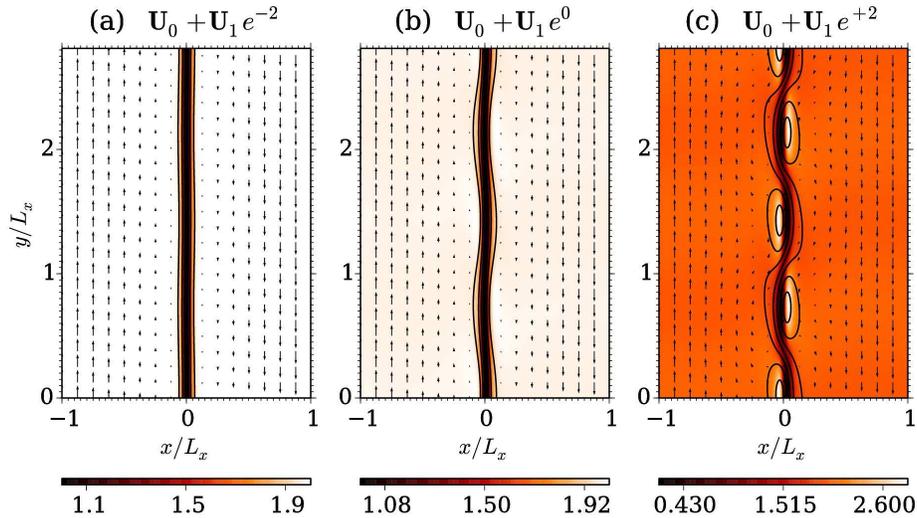}
   \caption{Unstable eigenfunction with $\beta=1$ and $k_yd=0.223$.
   The format is same as in Figure \ref{fig:3}.
   The growth rate is $0.675\Omega$.
   \label{fig:5}}
  \end{figure*}

  Figures \ref{fig:4} and \ref{fig:5} show the eigenfunctions for the
  cases with $\beta$=10 and 1, respectively, in the same format as in
  Figure \ref{fig:3}.
  The wavenumber is again assumed to be $k_yd$=0.223.
  Since the magnetic tension force, working as a restoring force,
  becomes stronger with an increase in the magnetic field strength, it
  becomes more and more difficult to bend the magnetic field lines
  significantly, and the unstable mode seems to be localized around the
  initial localized channel.
  It should be noted that the localization of the eigenfunction and its
  growth rate is a different matter.
  The growth rate actually becomes greater with the background magnetic
  field strength increased, as shown in Figure \ref{fig:1}, due to a
  large gradient of the field.
 
  This localization implies a small $B_{x1}$ compared with the
  initial $B_y$, but how it affects the average value of the Maxwell
  stress is not trivial.
  The estimate of the efficiency of the angular momentum transport is
  tightly connected with the non-linear behavior and the problem of the
  saturation mechanism.
  In the next section, we will discuss the significant contribution of
  the present instability to the stress tensor, using non-linear
  simulations.

 \section{NONLINEAR SIMULATIONS}\label{sec:nonlinear}
 This section provides the results of fully-non-linear MHD simulations
 designed to validate the presence of MGDIs suggested in
 the previous section and to investigate the non-linear time evolution.
 Specifically, we focus on the efficiency of angular momentum
 transport.

  \subsection{Basic Equations and Simulation Codes}
  We solve the same equations as used in the previous section,
  i.e., from equation (\ref{eq:linear1}) to (\ref{eq:linear4}), but
  rewritten in the semi-conservative form:
  \begin{equation}
   \frac{\partial \rho}{\partial t}
    + {\bf \nabla}\cdot\left(\rho {\bf v}\right) = 0,
  \end{equation}
  \begin{eqnarray}
   \frac{\partial \left(\rho {\bf v}\right)}{\partial t}
    + {\bf \nabla}\cdot
    \left[\rho{\bf vv}+\left(p+\frac{B^2}{2}\right){\bf I}
    -{\bf BB}\right]
   = - 2\rho{\bf \Omega}\times{\bf v}
    - 2\rho \Omega x v_{0y}^\prime \hat{\bf e}_x,
  \end{eqnarray}
  \begin{equation}
   \frac{\partial {\bf B}}{\partial t}
    = {\bf \nabla}\times\left({\bf v}\times{\bf B}\right),
  \end{equation}
  \begin{eqnarray}
   \frac{\partial e}{\partial t}
   +{\bf \nabla}\cdot
   \left[\left(e+p+\frac{B^2}{2}\right){\bf v}
    -\left({\bf v}\cdot{\bf B}\right){\bf B}\right]
   = - 2 \rho \Omega x v_{0y}^\prime v_x,
  \end{eqnarray}
  where $e=\rho v^2/2+p/(\gamma-1)+B^2/2$ is the total energy density,
  and the other notation is same as in the linear analysis.
  The specific heat ratio is again set to be $\gamma=5/3$.
  All quantities are spatially discretized by using the finite
  difference approach in a computational domain, $-1 \leq x/L_x \leq 1$
  and $0 \leq y/L_x \leq 2\pi$, which is resolved with 200$\times$600
  grid points.

  We calculate the flux with the help of the HLL approximate Riemann
  solver \citep{Harten1983} at the face center, where the primitive
  variables, i.e., $\rho$, ${\bf v}$, ${\bf B}$, and $p$, are 
  evaluated as point values by combining a 5th-order weighted
  essentially non-oscillatory (WENO) interpolation
  \citep{Liu1994,Jiang1996}
  and the monotonicity preserving limiter \citep{Suresh1997}.
  The point-value flux is then converted to the appropriate numerical
  flux with a 6th-order formula \citep{Shu1988}.
  The cell-centered conservative variables are finally updated using the
  3rd-order TVD Runge-Kutta method \citep{Shu1988}.
  To avoid a spurious magnetic monopole, the HLL-upwind constrained
  transport (UCT; \cite{Londrillo2004}) treatment is employed for
  updating the face-centered magnetic field, in which the edge-centered
  electric field is evaluated using WENO interpolation and the HLL
  average.
  Note that the results to be shown in this section will not
  largely change, even if the HLLD Riemann solver, which is more
  accurate than the HLL Riemann solver especially in high-$\beta$
  plasmas \citep{Mignone2007,Mignone2009}, is employed instead.
  The quantitative behavior of statistics, however, slightly differs due
  to the higher resolution of each wave mode.
  In particular, less diffusivity is preferable to larger stress related
  to turbulent motion in small scales.

  \subsection{A Single Localized B-Field}
  Our initial condition is set to be a superposition of exactly the same
  equilibrium state considered in our linear analysis and a random
  perturbation of the in-plane velocity, the amplitude of which is fixed
  to 1\% of the sound speed measured in an unmagnetized region.
  Without the random perturbation, the system would remain in the
  initial equilibrium state.
  To calculate the long time evolution of the system, we implement the
  standard shearing periodic boundary condition.
  Even if one adopts a conducting wall boundary, the results do not
  change in the early stage, before the distorted magnetic fields
  approach the radial boundary.

\begin{figure}[htbp]
   \epsscale{0.7}
   \plotone{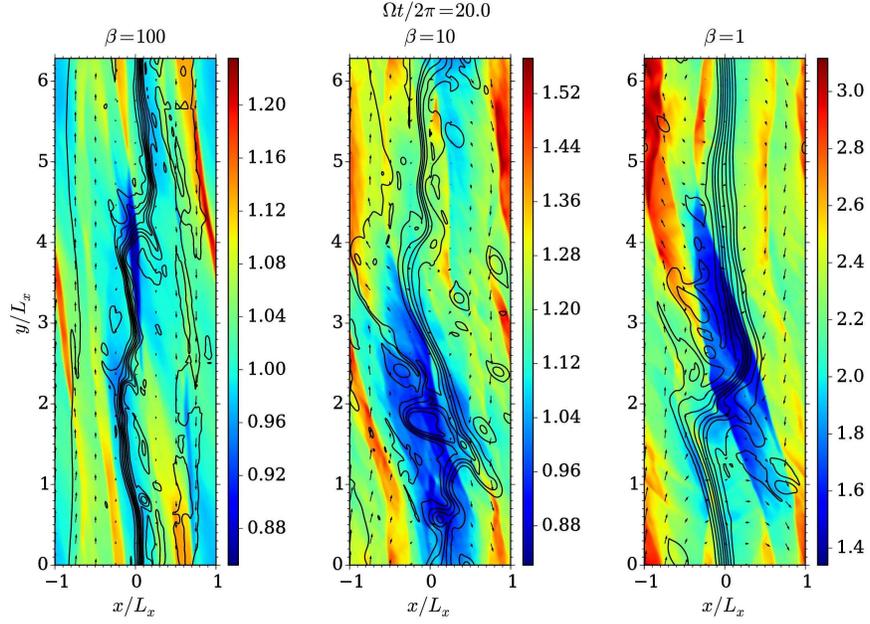}
   \caption{Snapshots of the MHD simulations at twenty times the orbital
   period.
   The color contours, solid lines, and arrows represent the gas
   pressure, the lines of magnetic force, and the in-plane velocity,
   respectively.
   \label{fig:6}}
  \end{figure}

  From left to right in Figure \ref{fig:6}, snapshots of the simulations 
  with $\beta$=100, 10, and 1 taken at time $\Omega t/2\pi=20$ are
  shown.
  The format of each panel is same as in Figures \ref{fig:3} to
  \ref{fig:5} except for the range of the $y$-coordinates.
  All of these cases show the negative correlation between $B_x$ and
  $B_y$ reflected as the downward-sloping magnetic field lines, which
  contributes to angular momentum transport.
  While the linear theory discussed in the previous section predicts
  broadband growth for a stronger initial field, the typical scale of
  the bending of the field line is clearly larger for smaller $\beta$,
  which suggests that the magnetic tension force works more efficiently
  in non-linear evolution and then the growth of short waves is
  suppressed.
  In larger $\beta$ cases, on the other hand, the bent mean structure
  and other small-scale magnetic structures appear.
  Such structures first grow along the both sides of the initial field,
  where the large gradient $\left|\partial B_y/\partial x\right|$
  exists, and then they are torn off from the mean field by magnetic
  reconnection.
  (Note that magnetic reconnection occurs via numerical resistivity, but
  the non-linear evolution does not change by assuming a finite
  resistivity.)
  In any case, the non-linear growth up to torsion of the localized
  magnetic field is ascertained.
  Note that the stage where the linear theory is applicable finishes
  instantly, since the growing perturbed field quickly breaks the
  background structure of the initial magnetic field.
  
\begin{figure}[htbp]
   \epsscale{0.7}
   \plotone{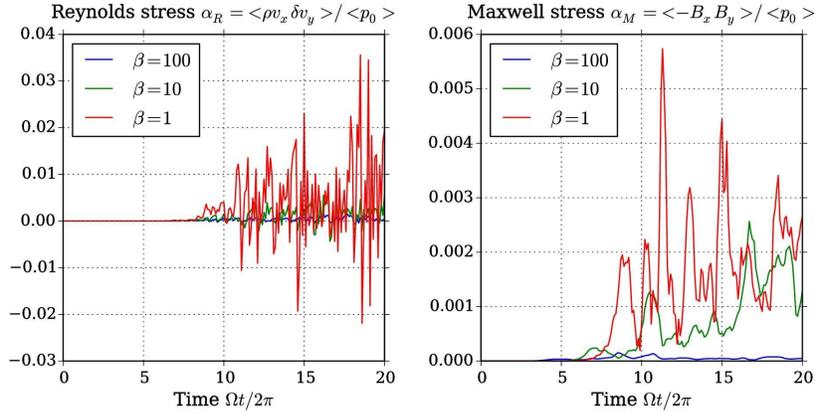}
   \caption{Time histories of the box-averaged stress.
   The left and right panels show the $xy$-components of the Reynolds
   stress and the Maxwell stress normalized by the averaged initial gas
   pressure, that is, the $\alpha$ parameters.
   \label{fig:7}} 
  \end{figure}

  Figure \ref{fig:7} summarizes the box-averaged stress as a function of
  time.
  The $xy$-components of the Reynolds stress and the Maxwell stress
  normalized by the initial gas pressure, or the so-called
  $\alpha$-parameters, are plotted in each panel.
  The instantaneous Reynolds stress generally fluctuates significantly
  with time, but the temporal average over the interval 
  $15 \le \Omega t/2\pi \le 20$ takes a positive value of the order of
  $10^{-3}$.
  The Maxwell stress, on the other hand, remains positive during the
  non-linear evolution, which is still smaller by one to two orders of
  magnitude compared with the local three-dimensional simulations of
  MRIs \citep[][etc.]{Hawley1995,Hawley1996}.
  Note that the case of $\beta=100$ shows a remarkably small value in
  spite of a more broken structure, due to the weakness of the initial
  field.
 
  To put it another way, however, once the magnetic field lines are
  stretched to the point of crossing the radial boundary, much stronger
  azimuthal fields could be expected.
  The point is a stretching term in the governing equation for the
  azimuthal magnetic energy,
  \begin{eqnarray}
   \frac{d}{dt}\left(\frac{B_y^2}{2}\right)
    = B_x B_y \frac{\partial v_y}{\partial x}
    - B_y^2 \frac{\partial v_y}{\partial x},
    \label{eq:omega-dynamo}
  \end{eqnarray}
  where the first and the second terms in the right-hand side represent
  the energy change by stretching and compressive motions,
  respectively.
  Especially, the energy increase through the background velocity shear,
  $B_x B_y v_{y0}^{\prime}$, is called the $\Omega$-dynamo.
  In a radially periodic system like the shearing box, if a fluid
  element moves largely across the radial boundaries, the total velocity
  shear the element feels can become much larger than the shear just
  inside one simulation domain, $q \Omega L_x$.
  The larger radial fluctuation, therefore, leads to the stronger
  azimuthal magnetic energy. 
  The next subsection is devoted to suggesting such a possible path
  leading to a more amplified magnetic field and a resultant large
  Maxwell stress.

  \subsection{Multiple Localized B-fields}
  As a phenomenon expected to occur in accretion disks, let us consider
  the situation where a toroidal magnetic field has multiple structures
  rather than a single localized field.
  The motivation of this idealized setup comes from, for example, the
  existence of parasitic instabilities occurring on a current sheet,
  which induce periodic variation of plasma parameters along an
  equatorial plane \citep{Goodman1994,Pessah2009,Rembiasz2016}.
  The physical mechanism driving the instability in the linear stage
  does not change even in this case, but a more turbulent state could be
  expected in the non-linear stage as a result of coupling between
  neighboring fields.
  This section shows a possible path through which the MGDIs
  may contribute to turbulence generation and anomalous angular momentum
  transport.

  The simulation setup is the same as described in the previous subsection
  except for the initial profile of the magnetic field.
  Here we assume the functional form of the toroidal magnetic field as
  follows:
  \begin{equation}
   B_{y0}\left(x\right) = B_c \cos^4\left(\frac{3\pi x}{L_x}\right),
  \end{equation}
  which reproduces both the locality within $0.05L_x$ and the
  periodicity of the localized structure.
  The pressure and density profiles are also modified to keep dynamical
  and thermal equilibrium.
  The plasma beta is defined using the peak value of the magnetic field,
  $B_c$, and the gas pressure at the same site.

\begin{figure}[htbp]
   \epsscale{0.7}
   \plotone{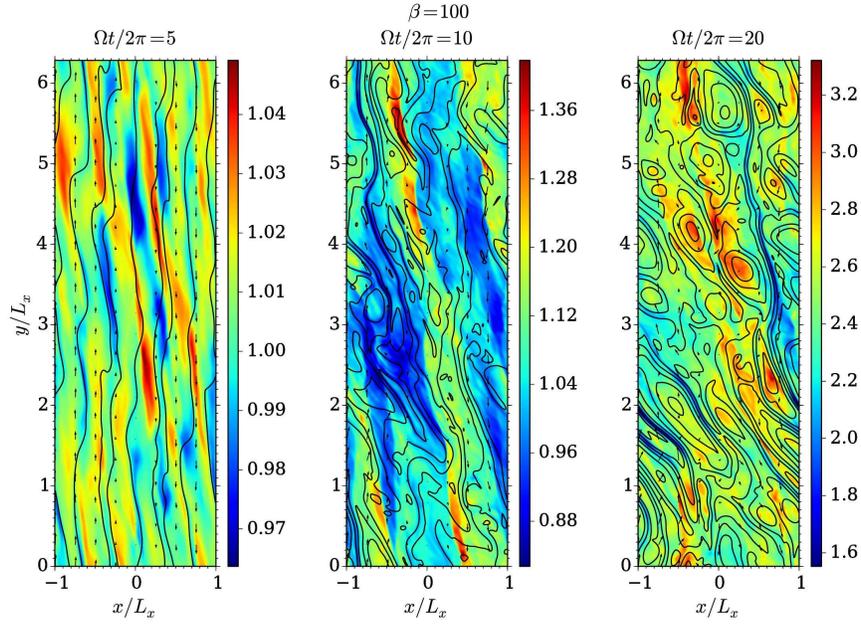}
   \caption{Snapshots of the simulation started from a wavy toroidal
   field with $\beta=100$, taken at 50, 100, and 200 orbits.
   The format of each panel is the same as in Figure \ref{fig:6}.
   \label{fig:8}}
  \end{figure}

\begin{figure}[htbp]
   \epsscale{0.7}
   \plotone{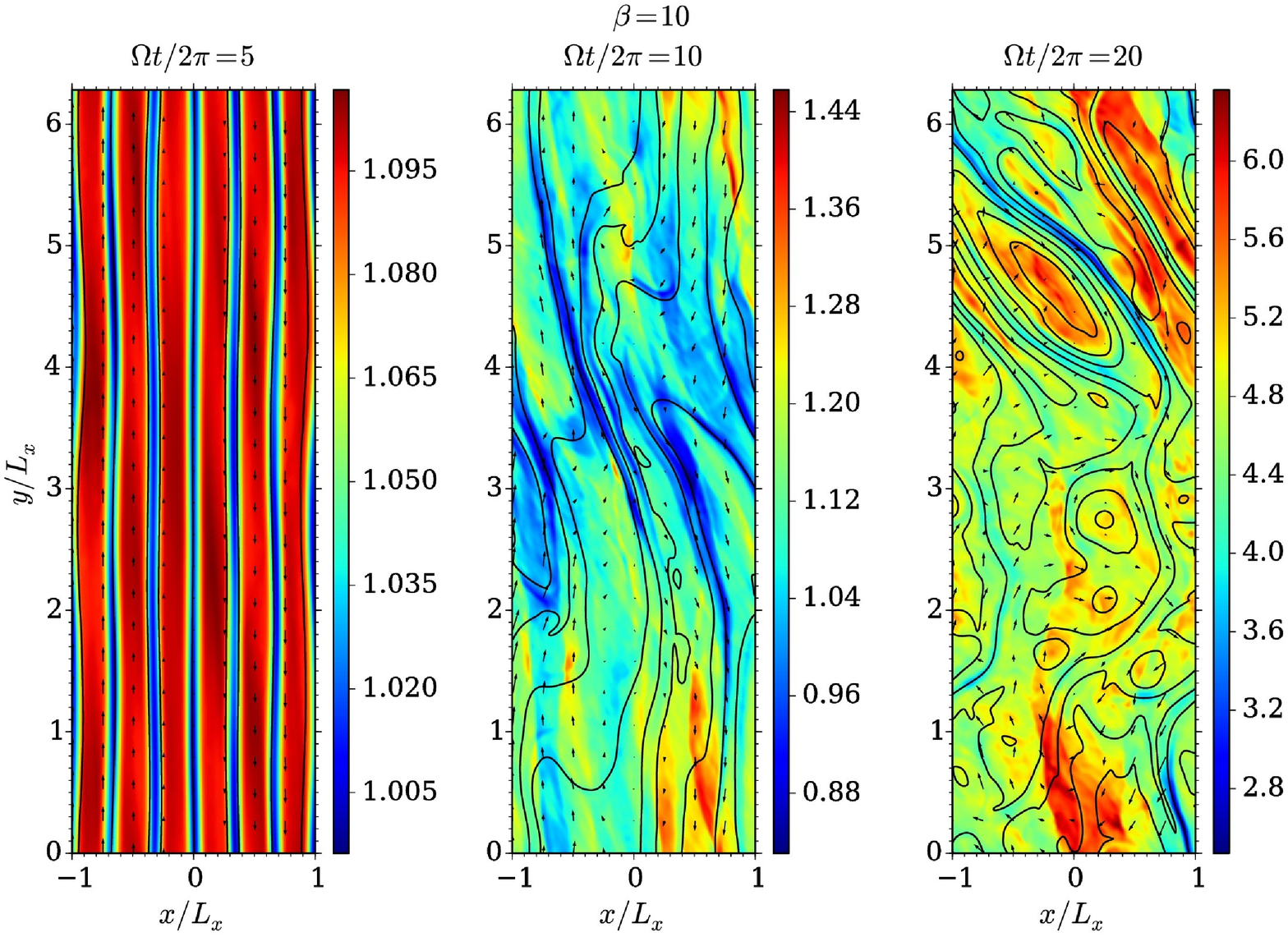}
   \caption{Snapshots of the case of $\beta=10$, with the same format as
   in Figure \ref{fig:8}.
   \label{fig:9}}
  \end{figure}

\begin{figure}[htbp]
   \epsscale{0.7}
   \plotone{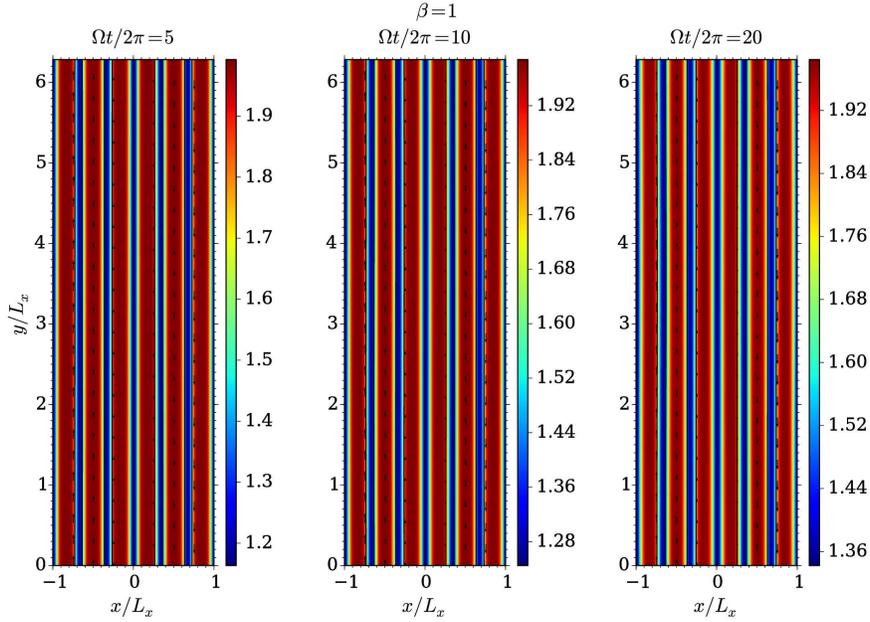}
   \caption{Snapshots of the case of $\beta=1$, with the same format as
   in Figure \ref{fig:8}.
   \label{fig:10}}
  \end{figure}

  Figures \ref{fig:8}, \ref{fig:9}, and \ref{fig:10} show snapshots at
  characteristic stages in the cases with $\beta$=100, 10, and 1,
  respectively.
  The three panels in each figure are taken at times $\Omega t/2\pi$=5,
  10, and 20 from left to right, and the format of each panel is the
  same as in Figure \ref{fig:6}.
  In the leftmost panel in Figure \ref{fig:8}, we can see that the
  discrete magnetic field lines are distorted individually in the early
  stage by the non-linear growth, just as demonstrated in the
  simulations with a single localized field.
  The bending of the field lines grows as time goes on, and before ten 
  orbital periods they drastically overlap and merge with the
  neighboring magnetic fields.
  The mixing of the magnetic field is completed by 20 orbital periods,
  and the simulation domain is filled with a lot of magnetic islands as
  a result of the repetitive reconnection process.
  Recall that, in a single channel case, the reconnected field simply
  gets torn off the background field and shows no further turbulent
  development.
  At this stage, the energy contained in the magnetic field increases to
  about 10\% of the kinetic energy of the background differential
  rotation.
  Even if the difference of the initial magnetic energy is taken into
  account, this ratio is rather large compared with the case of a single
  localized field, where the magnetic energy at the saturated stage is
  smaller by three orders of magnitude.
 
\begin{figure}[htbp]
   \epsscale{0.7}
   \plotone{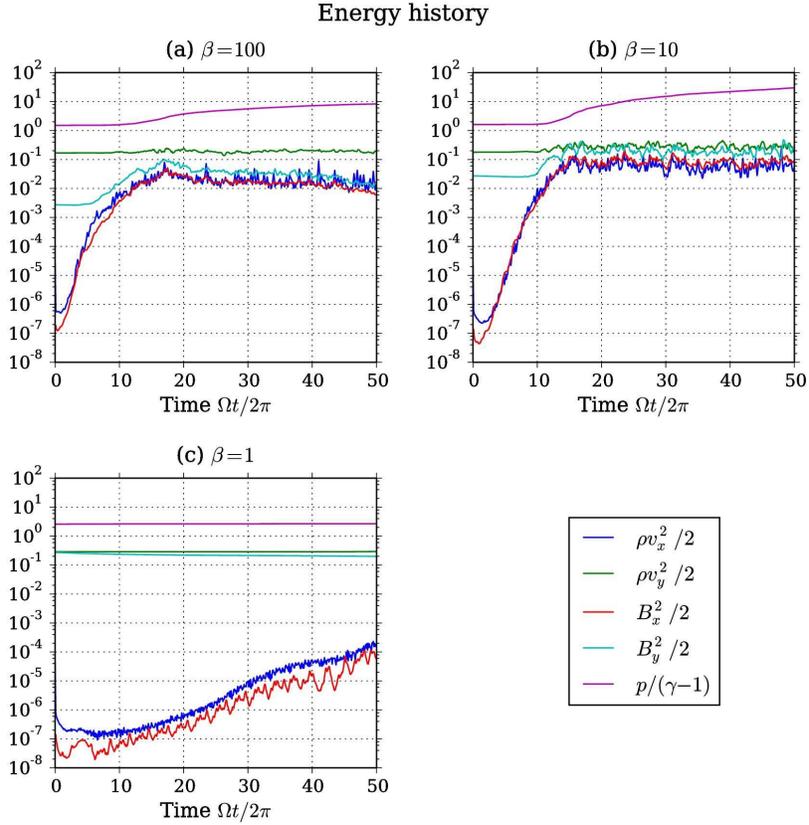}
   \caption{Temporal development of the box-averaged energy divided into
   the contribution from the $x$ and $y$ directions, all of which are
   normalized by the gas pressure measured in a magnetized region.
   Each panel shows the results for (a) $\beta=100$, (b) $\beta=10$, and
   (c) $\beta=1$, respectively.
   \label{fig:11}}
  \end{figure}

  The detailed time history of the box-averaged energy is summarized in
  Figure \ref{fig:11}(a), which shows that the dynamo process works
  efficiently on both $B_x$ and $B_y$ at an early stage before 20
  orbital periods, and roughly speaking, an equipartition is eventually
  attained between the kinetic energy in the $x$-direction and the
  magnetic energy in the $x$- and $y$-directions.
  The production of the strong azimuthal field is understood as a
  natural conseqence from equation (\ref{eq:omega-dynamo}), which
  tells the effect of $\Omega$-dynamo.
  A similar relation holds with regard to the radial field energy as
  \begin{eqnarray}
   \frac{d}{dt}\left(\frac{B_x^2}{2}\right)
    = B_x B_y \frac{\partial v_x}{\partial y}
    - B_x^2 \frac{\partial v_y}{\partial y},
  \end{eqnarray}
  which shows that shear motion in the radial velocity newly generates
  the radial magnetic field.
  It is clear that in the MGDI this radial velocity is driven by the
  magnetic pressure gradient force, while the gravity-related terms play
  the same role in the case of MRI.
  In this sense, the dynamo process working here looks quite similar to
  that in the MRI not only in the azimuthal field, but also in the
  radial field.
  At the saturated stage, the gas pressure gradually increases through
  magnetic reconnection, creating many magnetic islands.
 
  Even when the initial magnetic energy is ten times as large, that is,
  $\beta=10$, the same instability grows as shown in Figure
  \ref{fig:9}.
  Compared with the case of $\beta=100$, it can be seen that the typical
  scale of the growing mode becomes larger, as does the size of the
  magnetic islands in the final stage.
  The gas pressure at 20 orbital periods increases about twofold,
  which implies that the total amount of energy input into the system
  across the boundaries is enhanced and continuously converted to the
  internal energy of the plasma through magnetic reconnection.
  In other words, the total stress integrated along the radial boundary
  gets larger, since the time variation of the total energy, $e$, can be
  described as follows:
  
  \begin{eqnarray}
   \frac{\partial}{\partial t} \int e \, dV
    = 2 \left|v_{0y}(L_x)\right| \int_{x=L_x} W_{xy} dy
    - \int 2\rho\Omega x v_{0y}^\prime v_x dV,
    \label{eq:ev}
  \end{eqnarray}
  where $W_{xy}=\rho v_x \delta v_y - B_x B_y$ is the total stress.
  The second term on the right-hand side of equation (\ref{eq:ev}),
  which represents the change in the total gravitational potential in
  the simulation domain, becomes nearly zero on average.
  The detailed time history of each energy component is plotted in
  Figure \ref{fig:11}(b).
  As in the case of $\beta=100$ shown in panel (a), the growing mode
  is saturated before 20 orbital periods, after which the kinetic energy
  related to $v_x$ and the magnetic energy reach a level comparable to
  the background differential rotation, but slightly larger than those
  in the case with a weaker initial field.

  However, this situation changes significantly for the strong magnetic
  field with $\beta=1$ shown in Figure \ref{fig:10}.
  The initial equilibrium state holds and no growing mode can be
  observed.
  It is interesting to note that the non-linear evolution is
  accomplished in the single field case discussed in the previous
  subsection.
  The energy history in Figure \ref{fig:11}(c) also shows a quite calm
  variation.
  The suppression of the growing mode is indeed the result of the
  non-linear magnetic tension force, since we confirmed the presence of
  unstable modes in linear analyses under the initial magnetic profile
  described here.

\begin{figure}[htbp]
   \epsscale{0.7}
   \plotone{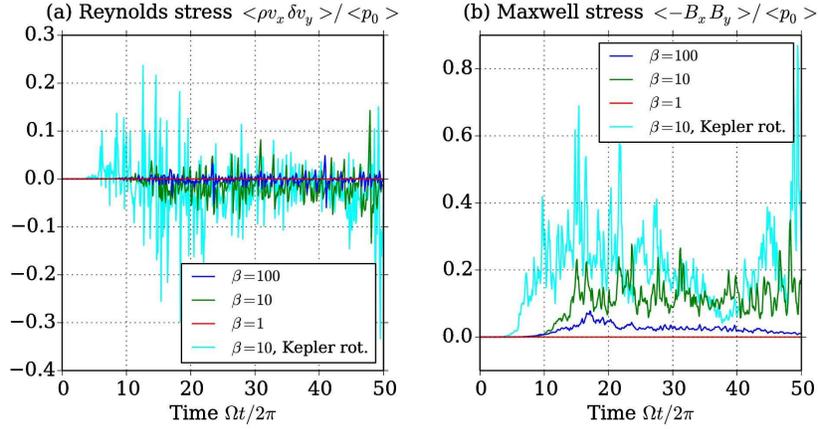}
   \caption{The $xy$-components of the (a) Reynolds and (b) Maxwell
   stress tensors as functions of time.
   \label{fig:12}}
   \end{figure}

  We summarize the time histories of the Reynolds and Maxwell stresses
  in Figures \ref{fig:12}(a) and \ref{fig:12}(b), respectively.
  The result of $\beta=1$, denoted by the red line, shows no stress
  for either the Reynolds or Maxwell components, because the initial
  equilibrium state is almost conserved.
  In the cases with $\beta=10$ and 100, denoted by the green and blue
  lines, the Reynolds and Maxwell stresses averaged after 20 orbital
  periods are -0.00109 and 0.0235 for $\beta=100$, and -0.0164 and 0.126
  for $\beta=10$, respectively.
  The Maxwell stress, therefore, is larger by about two orders of
  magnitude than the results in the single field case, which means a
  qualitative change in the non-linear behavior, rather than the simple
  quantitative superposition due to an increase in the initial total
  magnetic flux.
  In addition, the result for a stronger shear motion with $q=1.5$,
  which corresponds to the Kepler rotation, is also plotted as a cyan
  line.
  The basic mechanism driving the instability is same as in the case
  with $q=1.0$, but thanks to
  the more powerful $\Omega$-dynamo effect,
  the MGDI can grow more quickly non-linearly and a larger Maxwell
  stress can be attained at the saturated stage.

\begin{figure}[htbp]
   \epsscale{0.4}
   \plotone{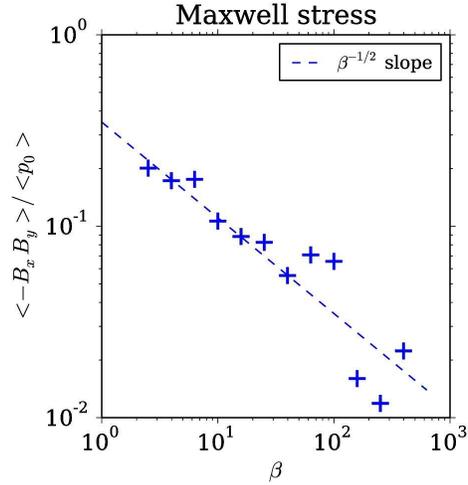}
   \caption{The Maxwell stress averaged during the saturated stage as a
   function of the initial $\beta$.
   \label{fig:13}}
  \end{figure}

  The $\beta$-dependence of the Maxwell stress averaged during the
  saturated stage between $30 \le \Omega t/2\pi \le 50$ is summarized in
  Figure \ref{fig:13}.
  It can be clearly seen that the results are well fitted by a power law
  of $\beta^{-1/2}$ as long as $\beta>2$, which indicates the
  proportionality to the initial magnetic flux rather than magnetic
  energy density, while the cases starting with $\beta<2$ result in
  almost no stress.

\begin{figure}[htbp]
   \epsscale{0.7}
   \plotone{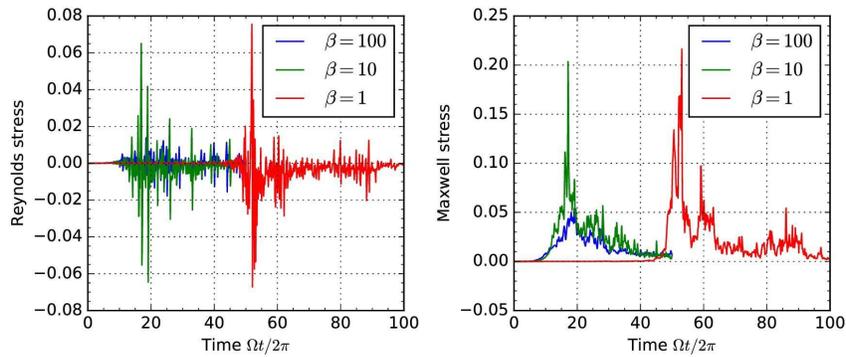}
   \caption{Time histories of the Reynolds and Maxwell stress normalized
   box averages of the instantaneous gas pressure in the elongated
   simulation box, $(x,y)\in(2L_x, 4\pi L_x)$, for defferent initial
   $\beta$.}
   \label{fig:longY}
  \end{figure}

  The suppression of instabilities for a small $\beta$ is highly
  relevant to the large characteristic spatial scale under the strong
  magnetic tension force, which works less efficiently for longer
  waves.
  Although this mode has a broad unstable range in wavelength with
  respect to the linear theory around a single channel, the nonlienar
  growth is actually quenched in the mutiple channel case.
  Once the simulation box is extended double in the azimuthal direction,
  however, the drastic growth in the Maxwell stress via channel merging
  process is activated even for $\beta=1$. 
  This certainly heppens because the extended box allows the growth of
  the fluctuations with the scale larger than the original domain size.
  Figure \ref{fig:longY} shows the time histories of the Reynolds and
  Maxwell stress for $\beta$=1, 10, and 100, respectively, normalized by
  the volume averages of the instantaneous thermal pressure,
  $\left<p\left(t\right)\right>$.
  From the right panel, we can clearly observe the enhancement of the
  Maxwell stress by the nonlienar growth after 50 orbits for $\beta=1$.
  Except for the shifted time this drastic merging is switched on, the
  statistical behavior is quite similar in all cases, which implies that
  the MGDI works as the common underlying mechanism to drive the
  turbulence.

 \section{SUMMARY AND CONCLUSION}\label{sec:summary}
 In the present paper, we have proposed a new plasma instability, MGDI,
 which can generate a highly turbulent state in an accretion disk and
 contribute to the enhanced transport of angular momentum.
 Driven by the spatial non-uniformity of a toroidal magnetic field, an
 unstable mode completely confined within the equatorial plane can be
 realized, in contrast to the previous studies on the toroidal MRI, in
 which a vertical wavenumber always dominates over a finite azimuthal
 component.
 
 The growth rates and eigenfunctions of this instability are calculated
 by linear eigenvalue analysis, and the corresponding non-linear
 evolution is then demonstrated using two-dimensional MHD simulations.
 While the simulations beginning with a single localized toroidal
 field reveal the non-linear growth of the MGDIs, it is shown that the
 angular momentum transport does not work so efficiently in the
 saturated stage.
 If the instabilities occur in the neighboring field lines under the
 multiple localized magnetic fields rather than in an isolated
 situation, however, they drastically overlap with each other and a
 well-developed turbulent state can be realized.
 In such a case dynamo action by differential rotation begins to work
 efficiently on magnetic field lines crossing the radial boundaries,
 which contributes to a large Maxwell stress.
 Furthermore, we have shown that a toroidal field with a larger
 magnetic flux is favorable for the Maxwell stress to reach a large
 value, but this drastic transition does not occur for a magnetic field
 that is too strong with $\beta<2$,
 as long as we employ the box size $(x,y) \in (2L_x, 2\pi L_x)$.
 Once the simulation domain is elongated, the transition is reactivated,
 but it takes much longer time to switch on the drastic merging.
 It is worth noting that the situation with $1 \le \beta \le 10$, which
 is favorable to the growth of the MGDI, often appears during a
 non-linear phase in a local simulation of the MRI for a relatively
 small initial beta, $\beta\sim\mathcal{O}\left(10^2\right)$.
 In the cases with larger initial $\beta$ values, like
 $\beta\sim\mathcal{O}\left(10^{3-6}\right)$,
 the final states still seem to be in the unstable regime
 \citep[e.g.,][]{Hawley1995,Sharma2006,Minoshima2015}.

 Although the profile of the toroidal magnetic field discussed here is
 one of the most idealized situations, the physical essence to drive the
 instability does not change even under a different structure, as long
 as an enough radial gradient of the magnetic field is available.
 There is, therefore, possibility that the present unstable mode will
 arise around various kinds of fluctuation in the toroidal field, like
 via parasitic modes including the Kelvin-Helmholtz instabitlity and the
 tearing instability.
 
 It should be emphasized again that the efficiency of the anglar
 momentum transport obtained here is comparable to that obtained in
 evaluation in three-dimensional simulations of MRIs assuming an
 initial toroidal field, in spite of the low dimensionality, and
 therefore, the MGDIs are capable of driving strong turbulence
 alone.
 One might expect not just the sole contribution, but also coupling
 with magnetic reconnection occurring parallel to the equatorial plane
 during the nonlinear phase of MRI, and with the toroidal MRIs as
 considered in previous studies if vertical variation is also taken into
 account.
 It is still not obvious that how large contribution the instabilities
 have in fully three-dimensional shearing boxes.
 Nevertheless, since they provide new paths toward turbulence without
 any finite $k_z$ in contrast to the conventional toroidal MRIs, the
 complementary growth between toroidally and vertically propagating
 waves, rather than competitive growth, should be expected.
 The present instability could possess the ability to play a wide
 variety of crucial roles in the mechanism of turbulence generation in
 differentially rotating systems.

 \acknowledgments
 We thank T. Amano and T. Saito for their useful discussion.
 This work was supported by JSPS KAKENHI Grant Number 26$\cdot$394.

\end{document}